\documentclass[aps,prd,preprintnumbers,nofootinbib,preprint]{revtex4}
\usepackage{bm}
\usepackage{latexsym}
\usepackage{dcolumn}
\usepackage{amsmath,amsfonts,amssymb}
\usepackage{graphicx,epsfig}
\usepackage{amsmath,amssymb}

\def\be {\begin{equation}}
\def\ee {\end{equation}}
\def\bea {\begin{eqnarray}}
\def\eea {\end{eqnarray}}
\def\bc {\begin{center}}
\def\ec {\end{center}}
\def\bfg {\begin{figure}}
\def\efg {\end{figure}}
\def\bi {\begin{itemize}}
\def\ei {\end{itemize}}
\def\nn {\nonumber}

\begin{document}

\title{The covariant, time-dependent Aharonov-Bohm Effect}

\author{Douglas Singleton $^{1,2}$} \email[email: ]{dougs@csufresno.edu}
\author{Elias C. Vagenas $^3$} \email[email: ]{evagenas@academyofathens.gr}

\affiliation{$^1$~Physics Department, CSU Fresno, Fresno, CA
93740-8031, USA}

\affiliation{$^2$~Department of Physics, Institut Teknologi
Bandung, Bandung, Indonesia}

\affiliation{$^3$~Research Center for Astronomy and Applied
Mathematics, Academy of Athens, Soranou Efessiou 4, GR-11527,
Athens, Greece}


\begin{abstract}
\par\noindent
We discuss two possible covariant generalizations of the
Aharonov-Bohm effect - one expression in terms of the space-time line integral
of the four-vector potential and the other expression in terms of the
space-time ``area" integral of the electric and magnetic fields written in
terms of the Faraday 2-form. These expressions allow one to calculate the
Aharonov-Bohm effect for time-dependent situations. In particular, we use these expressions to
study the case of an infinite solenoid with a time varying flux and find that the
phase shift is zero due to a cancellation of the Aharonov-Bohm phase shift with
a phase shift coming from the Lorentz force associated with the electric
field, ${\bf E} = - \partial _t {\bf A}$, outside the solenoid. This result may
already have been confirmed experimentally.
\end{abstract}

\maketitle
%
%
\par\noindent {\bf 1. Introduction} \\
%
%
\par
The Aharonov-Bohm (AB) effect \cite{AB,ES} lies
at the interface of gauge theories and quantum mechanics. In its
best known form, the AB effect predicts a shift in the interference
pattern of the quantum mechanical double-slit experiment which has
a magnetic flux carrying solenoid placed between the slits. If a
solenoid with a magnetic field ${\bf B} = \nabla \times {\bf A}$
(where ${\bf A}$ is the electromagnetic vector potential) is
placed between the two slits of a double-slit experiment the phase,
$\alpha$,  of the wave-function of the electrons going through the
slits and following some $path$ to the screen will be shifted by
an amount
\begin{equation}
\label{phase-a} \alpha _B= -\frac{e}{\hbar} \int _{path}
{\bf A} \cdot d{\bf x}
\end{equation}
where $e$ is the charge of the electron. If one
considers two electrons arriving at the screen via two separate
path, namely $path_1$ and $path_2$, one can reverse one of the paths and
find that the phase difference between the two electrons at the
screen is given by
\begin{equation}
\label{phase-a1}
\delta \alpha _B = \alpha_{B_1} - \alpha _{B_2}
 =  \frac{e}{\hbar} \oint _{2 - 1} {\bf A}
\cdot d {\bf x} =  \frac{e}{\hbar} \int {\bf B} \cdot d{\bf S} =
 \frac{e}{\hbar} \Phi _0
\end{equation}
where the subscript $2-1$ means the path going from the slits to
the screen along $path_2$ and returning along $path_1$. We used
Stokes' theorem on the closed line integral and $\nabla \times {\bf
A} = {\bf B}$. Finally, $\Phi _0$ is the magnetic flux through the
cross sectional area, ${\bf S}$, of the solenoid. This shift in the
phase leads to a shift in the position, $x$, of the interference
pattern maxima and minima on the screen by $\Delta x = \frac{L
\lambda  }{ 2 \pi d}\delta \alpha _B$ where $L$ is the distance to
the screen, $d$ is the distance between the slits and $\lambda$ is
the wavelength of the wave-function. This shift due to the
magnetic AB effect has been measured \cite{chambers,tonomura}.
Note, there is some unavoidable arbitrariness in
the sign $\delta \alpha _B$ depending on the rotational sense of
the closed loop going around the solenoid -- going along $path_1$
to the screen and returning along $path _2$ versus going along
$path_2$ to the screen and returning along $path_1$. However, the
shift of the interference pattern is independent of this
arbitrariness.

The importance of the magnetic AB effect of \eqref{phase-a1} is
that it shows (to some degree) the physical nature of the vector
potential, ${\bf A}$, since the electrons move in a region,
outside the solenoid, where ${\bf B}=0$ but ${\bf A} \ne 0$.
However, although ${\bf A}$ is gauge {\it variant} under the gauge
transformation ${\bf A} \rightarrow {\bf A} - \nabla \Lambda$,
where $\Lambda ({\bf x}, t)$ is some arbitrary function, the final
result in the phase difference, $\delta \alpha _B$, is gauge independent
since it can be turned into a surface integral of the magnetic
field, which is gauge invariant.

The electric version of the AB effect has been less discussed and
investigated. It was experimentally observed relatively recently in
\cite{AB-electric}. Similar to the magnetic case above, one can
show \cite{AB} that for an electron moving through some
region of space with an electric scalar potential $\phi$, it will
have its phase shifted by an amount
\begin{equation}
\label{phase-p} \alpha _E = \frac{e}{\hbar} \int ^{t_2}
_{t_1} \phi dt
\end{equation}
where $\Delta t = t_2 -t_1$ is the time the electron spends in the
potential. If one considers electrons moving along two different
paths, $path_1$ and $path_2$, with different values of the
potential, $\phi_1$ and $\phi_2$, along the different paths, then
the electrons will acquire a phase difference due to traveling in
different potentials given by
\begin{equation}
\label{phase-p1} \delta \alpha _E = \frac{e}{\hbar} \int ^{t_2}
_{t_1} \Delta \phi dt  = \frac{e}{\hbar} \int ^{t_2} _{t_1} \int
{\bf E} \cdot d{\bf x} ~ dt
\end{equation}
where $\Delta \phi = \phi _2 - \phi _1 = - \int ^1 _2 \nabla \phi
\cdot d{\bf x} = \int {\bf E} \cdot d {\bf x}$ is the potential
difference between the two paths through which the electrons move.
The last form of the electric phase shift in \eqref{phase-p1}, i.e.
$\frac{e}{\hbar} \int ^{t_2} _{t_1} \int {\bf E} \cdot d{\bf x} ~ dt$,
appears similar to the last form of the magnetic phase shift in
\eqref{phase-a1}, i.e. $\frac{e}{\hbar} \int {\bf B} \cdot d{\bf S}$, in
that both have the form $\delta(Phase) \propto (Field) \times (Area)$
although for the electric phase shift the ``area" has one space side and one
time side while the magnetic phase shift has a conventional area having two
space sides. One can flesh out this connection via the following
heuristic argument: For a small distance, $\Delta {\bf x}$, between
the two different potentials, $\phi_2$ and $\phi _1$, one can write
$\Delta \phi = {\bf E} \cdot \Delta {\bf x}$. Using this in the
first expression in \eqref{phase-p1}, one can write $\delta \alpha
_E = \frac{e}{\hbar}({\bf E} \cdot \Delta {\bf x}) \Delta t$,
where again $\Delta t$ is the time that the two electrons spend in
their respective potentials. Now $\Delta t \propto L$ where $L$ is
the length of the region through which the electrons move where
the potentials are $\phi_2$ and $\phi _1$ - more precisely $v_e
\Delta t  = L$ where $v_e$ is the speed of the electrons as they
move through these regions of constant scalar potential. Combining
these results, we find $\delta \alpha _E \propto ({\bf E} \cdot
\Delta {\bf x}) L$, and $(\Delta {\bf x}) L$ is the area between
the two tubes of length $L$ separated by a distance $\Delta {\bf
x}$, i.e. $d( Area) = (\Delta {\bf x}) L$. Thus, both magnetic and
electric AB phase differences from \eqref{phase-a1} and \eqref{phase-p1} can
be written in the form $\delta(Phase) \propto (Field) \times
(Area)$. Pictorially, one can see this $(Area)$ as the area swept out
by an imaginary string which connects the two electrons - the
length of the string is $\Delta {\bf x}$ and the length swept out
is $L$. Note that the phase difference in \eqref{phase-p1} is in
addition to any phase difference due to the path length difference
between $path_1$ and $path_2$. Also, as in the magnetic case
\eqref{phase-a1}, there is an unavoidable sign ambiguity in
\eqref{phase-p1} depending whether one considers $\Delta {\bf
x}$ as coming from a path going from $\phi _1$ to $\phi_2$ or,
alternatively, a reversed path going from $\phi _2$ to $\phi_1$.

The expressions \eqref{phase-a1} and \eqref{phase-p1} are
written in three-vector form so they are not obviously covariant.
In the next section, we examine two possible covariant generalizations
of the AB phase differences \eqref{phase-a1} and \eqref{phase-p1} which allow
one to examine time dependent Aharonov-Bohm experiments. \\\
%
%
\par\noindent {\bf 2. Covariant expressions for the AB phase shift} \\
%
%
\par
The first covariant version of the AB phase differences generalizes the
potential form of the phase difference given by the first expressions
on the right hand side of \eqref{phase-a1} and \eqref{phase-p1}
\begin{equation}
\label{cov-1} \delta \alpha _{EB} = \frac{e}{\hbar} \oint A_{\mu}
dx^{\mu} = \frac{e}{\hbar} \left[ \int ^{t_2} _{t_1} \Delta \phi dt -
\oint {\bf A} \cdot d{\bf x} \right] ~.
\end{equation}
This covariant expression for the AB phase shift was used in \cite{AB}.
The closed loop integral in the four-vector expression,
$\oint A_{\mu} dx^{\mu}$, is not a closed time loop but is to
be taken in the sense that the two electron both start at
the space-time point $(t_i, {\bf x}_i)$, travel along
two different paths, $path_1$ and $path_2$, and end up at the same
space-time point $(t_f, {\bf x}_f)$ with $t_f > t_i$. One reverses
the direction of one of the paths and in this way gets $\Delta \phi = \phi _2 - \phi _1$
in the time integral and one gets a closed loop for the spatial
integral, i.e. $\oint {\bf A} \cdot d{\bf x}$.

The second covariant version of the AB phase difference
generalizes the $(Field) \times (Area$) form of the
phase difference, i.e. the last two expressions for the magnetic
and electric phase differences given in \eqref{phase-a1} and \eqref{phase-p1}.
This second covariant version of the AB phase is best expressed in the
notation of differential forms and the wedge product \footnote{For our purposes the
elementary and excellent introduction to differential forms given in \cite{ryder}
is all we will need.}. This second proposed expression for the covariant AB phase is
\begin{equation}
\label{cov-2} \delta \alpha _{EB} = -\frac{e}{2 \hbar} \int F_{\mu
\nu} dx^\mu \wedge dx^\nu = \frac{e}{\hbar} \int F
\end{equation}
where $F= -\frac{1}{2} F_{\mu \nu} dx^\mu \wedge dx^\nu$ is the
Faraday 2-form, $dx^\mu$ and $dx^\nu$ are differential four-vectors, and
$\wedge$ is the antisymmetric wedge product \cite{ryder}. The
factor of $\frac{1}{2}$ accounts for the anti-symmetry of $F_{\mu
\nu}$ and $dx^\mu \wedge dx^\nu$.

We now expand the Faraday 2-form out, and show that is
reproduces the standard, static AB phase results \eqref{phase-a1}
and \eqref{phase-p1},
\begin{eqnarray}
\label{faraday}
F &=& -\frac{1}{2} F_{\mu \nu} dx^\mu \wedge dx^\nu \nn \\
&=& (E_x dx+ E_y dy + E_z dz) \wedge dt
+ B_x dy \wedge dz + B_y dz \wedge dx + B_z dx \wedge dy ~.
\end{eqnarray}
If the electric field is zero, i.e. ${\bf E} =0$, then one has $F = B_x dy
\wedge dz + B_y dz \wedge dx + B_z dx \wedge dy = {\bf B} \cdot
d{\bf S}$ where the differential forms expression has been
converted back to three-vector notation and $d{\bf S}$ is the
differential area. Thus, the expression in \eqref{cov-2} reduces to
$\delta \alpha _{EB} = \frac{e}{\hbar} \int F = \frac{e}{\hbar}
\int {\bf B} \cdot d{\bf S}$ which is equivalent to the
three-vector expression \eqref{phase-a1}.

If, on the other hand, the magnetic field is zero, i.e. ${\bf B} =0$,
{\it and} one has a time independent system (so that $\partial _t
{\bf A} =0$ and ${\bf E} = - \nabla \phi$), then the non-zero
terms of the Faraday 2-form are
$F = -\partial_x \phi dx \wedge dt - \partial_y \phi dy \wedge dt -
\partial _z \phi dz \wedge dt$. Doing the spatial integral of
this expression for the Faraday 2-form yields $-\int ^1 _2 \nabla
\phi \cdot d{\bf x} = \phi _2 - \phi _1 =\Delta \phi$. Thus, under
these conditions, the expression in \eqref{cov-2} reduces to
$\delta \alpha _{EB} = \frac{e}{\hbar} \int F = \frac{e}{\hbar}
\int \Delta \phi dt$ which is equivalent to the first three-vector
expression on the right hand side of \eqref{phase-p1}.

In summary, in this section, we have constructed two covariant
versions of the AB phase difference, \eqref{cov-1} and \eqref{cov-2}.
In the next section, we will discuss how one can experimentally
test these covariant expressions for the AB phase difference,
\eqref{cov-1} or \eqref{cov-2}, in the time dependent situation of an infinite
solenoid with a time varying magnetic flux. \\
%
%
\par\noindent {\bf 3. Solenoid with time varying flux} \\
%
%
\par
For static situations, both \eqref{cov-1} and \eqref{cov-2}
reproduce the results for the magnetic and electric AB phase differences
\eqref{phase-a1} and \eqref{phase-p1}. However, for certain time
dependent situations, the two expressions both lead to the conclusion that
there is an exact cancellation of the magnetic and electric AB
phase shifts so that one finds no net phase shift differences coming from the
time-dependent electromagnetic field.
In particular, we have in mind the usual magnetic AB set-up of an infinite solenoid
but with a time dependent magnetic field and vector potential, i.e. ${\bf B} (t)$, and
${\bf A} (t)$. Note that for this situation the scalar potential is still
zero, $\phi =0$. At first, one might think that for this set-up the AB phase
would simply be obtained by inserting ${\bf A} (t)$ into the first
expression on the right hand side of \eqref{phase-a1}, or inserting ${\bf B} (t)$
into the second expression on the right hand side of \eqref{phase-a1},
giving the usual magnetic AB phase shift \eqref{phase-a1} but with the
time dependence of the vector potential, i.e. $\delta \alpha _B \propto \Phi _0 (t)$. This is
in fact what previous work \cite{chiao, ageev, chentsov}
on the time dependent AB effect has suggested -- that there would be a time dependent
AB phase. However, for this time dependent set-up one can see there are
complications since unlike the static solenoid set-up, there is now a non-vanishing electric field
outside the solenoid coming from ${\bf E} = - \partial _t {\bf A}$. This induces an
additional phase shift as we will show below.

We will first calculate the AB phase difference predicted by \eqref{cov-2}. The part of the
AB phase difference from the three magnetic field terms of \eqref{faraday} is
\begin{equation}
\label{AB-B}
\frac{e}{\hbar} \int \left[ B_x dy \wedge dz + B_y dz
\wedge dx + B_z dx \wedge dy \right]
= \frac{e}{\hbar} \int {\bf B} ({\bf x} , t ) \cdot d{\bf S}
\end{equation}
where in the last expression we have converted back to
three-vector notation. The contribution to the covariant AB phase difference from the
three electric field terms of \eqref{faraday} is
\begin{equation}
\label{AB-E}
\frac{e}{\hbar} \int \left[ (E_x dx+ E_y dy + E_z dz)
\wedge dt \right] = - \frac{e}{\hbar} \oint {\bf A} \cdot d{\bf x}
= -\frac{e}{\hbar} \int {\bf B} ({\bf x} , t ) \cdot d{\bf S}
\end{equation}
where we have taken into account that ${\bf E} = - \partial _t {\bf A}$, performed
the $dt$ integration, and in the last expression we have used Stokes'
theorem. The $dt$ integration in \eqref{AB-E} has turned
${\bf E} = -\partial_t {\bf A}$ into $-{\bf A}$. The magnetic
contribution from \eqref{AB-B} is the negative of the electric
contribution from \eqref{AB-E} and the two parts cancel exactly.

We now calculate the phase shift for the time dependent,
infinite solenoid using \eqref{cov-1} for an infinitesimal arc. First,
the vector potential outside an infinite solenoid which has a time dependent
magnetic field and, therefore, a time-dependent current, $I(t)$, is
\begin{equation}
\label{A}
{\bf A} =\frac{k I(t)}{r} {\hat \theta}
\end{equation}
where $k$ is a constant whose exact form is not important for the present
and ${\hat \theta}$ is a unit vector in the angular direction.
Without loss of generality, we take the infinitesimal path
of the particle of charge $e$ and mass $m$ to be along
a circular arc, i.e. $d{\bf x} \propto {\hat \theta}$. Since
${\bf A} \propto {\hat \theta}$, the product ${\bf A} \cdot d{\bf x}$ will pick out
the angular direction of $d{\bf x}$. The relationship between
the angular displacement of the particle, $\Delta \theta$, the radius of the
arc, $r$, and the velocity of the particle, $v$, is
\begin{equation}
\label{dtr}
r \Delta \theta  = v \Delta t ~.
\end{equation}
Actually since the particle is accelerated by the electric field outside the
solenoid one should use $v \rightarrow (v_f + v_i)/2$, i.e. the average velocity
using the mid-point (this assumes that the acceleration due to the electric
field is a constant during this infinitesimal interval). Evaluating
$\frac{e}{\hbar} \int {\bf A} \cdot d {\bf x}$ for this infinitesimal path
, $d {\bf x} = r \Delta \theta {\hat \theta}$ , gives
\begin{equation}
\label{Aint}
\frac{e}{\hbar} \frac{k I(t)}{r} (r \Delta \theta) = \frac{e k I(t) \Delta \theta}{\hbar} ~.
\end{equation}
The question that arises now is ``Where does one evaluate $I(t)$?"; ``At the
initial time $t_i$, or final time $t_f$?". Based on the fact that we use
$v \rightarrow (v_f + v_i)/2$ for the velocity, we
evaluate $I(t)$ at the midpoint time $t= t_i + \Delta t/2$. Inserting
this into $I(t)$ and expanding to first order gives
$I(t_i + \Delta t/2) \approx I(t_i) + I'(t_i) \frac{\Delta t}{2} + ...$, with the
prime indicating a time derivative.
The first term, $I(t_i)$, is a constant and represents the initial,
static AB phase contribution. We can, without loss of generality,
take the initial current to be zero, $I(t_i) =0$, so that
there will be no initial phase shift. If there were a non-zero initial current,
one would instead have a constant phase shift of
$\delta \alpha _B =  \frac{e k I(t_i) \Delta \theta}{\hbar}$. Next, inserting
the second term in the expansion into \eqref{Aint} gives,
\begin{equation}
\label{Aint-2}
\delta \alpha_{A(t)} = \frac{e k I'(t) \Delta t \Delta \theta}{2 \hbar}
\end{equation}
which is the phase shift due to the time-dependence of the vector potential.
However, \eqref{Aint-2} is not the total phase shift in this case since there will
be an electric field, ${\bf E} = - \partial _t {\bf A}$, outside the solenoid which will
also contribute to the phase shift. We now calculate this shift. The electric field outside the solenoid is
\begin{equation}
\label{E}
{\bf E} = -\frac{\partial {\bf A}}{\partial t} = - \frac{k I'(t)}{r} {\hat \theta} ~.
\end{equation}
The acceleration associated with this electric field for the particle is
\begin{equation}
\label{acc}
{\bf a} =\frac{e {\bf E}}{m} = - \frac{e k I'(t)}{m r} {\hat \theta} ~.
\end{equation}
The change in distance, $\Delta d$,  due to the acceleration in \eqref{acc} is
\begin{equation}
\Delta d = \frac{1}{2} a \Delta t ^2 = - \frac{e k I'(t)}{2 m r} \frac{r \Delta \theta}{v} \Delta t
= - \frac{e k I'(t) \Delta \theta \Delta t}{2 m v} ~,
\end{equation}
where we have taken the odd (but perfectly legal) step of writing one of the $\Delta t$ factors as
$r \Delta \theta / v$ - see equation \eqref{dtr}. The change in
phase, $\delta \alpha_{E-field}$, due to this change in distance, $\Delta d$,
coming from the acceleration due to the electric field in \eqref{E} is
just $\Delta d$ divided by $\frac{\lambda}{2\pi}$ where $\lambda$ is the
de Broglie wavelength of the particle, i.e. $\lambda = \frac{h}{mv}$. Putting all these
together gives the phase shift due to the electric field as
\begin{equation}
\label{ephase}
\delta \alpha_{E-field} = \frac{\Delta d}{\lambda / (2 \pi)} = - \frac{e k I'(t) \Delta t \Delta \theta}{2 \hbar}~.
\end{equation}
One can see that the AB phase shift due the time variation of the potential given in \eqref{Aint-2} is canceled
exactly by the phase shift due to the effect of the electric field given in \eqref{ephase}, i.e.
$\delta \alpha _{A(t)} + \delta \alpha_{E-field} =0$. This leaves only the phase shift due to
any initial, static current and magnetic flux.

Thus, both versions of the covariant AB phase, \eqref{cov-1} and \eqref{cov-2},
predict that there will be no time-dependent AB phase shift for the solenoid
with a time dependent current and magnetic flux. For the covariant
phase shift expression in terms of the four-potentials \eqref{cov-1}
this result comes from a cancellation
between an AB type phase shift due to the time variation of
the vector potential \eqref{Aint-2} and a phase
shift due to the electric field \eqref{ephase}. For the covariant
phase shift expression \eqref{cov-2}, this result comes from an equivalent
cancellation between the electric contribution (the first three
terms of the Faraday 2-form) and the magnetic
contribution (the last three terms of the Faraday 2-form).

Previous works on this problem of the time dependent AB effect \cite{chiao, ageev, chentsov}
predicted that one should see a time dependent phase shift for a time dependent
vector potential and magnetic field. This is in contrast to the prediction
from the covariant expressions \eqref{cov-1} or \eqref{cov-2} that there will be no time dependent phase
shift due to a cancellation between the magnetic and electric parts of these expressions.
The suggested \cite{chiao, ageev} and performed \cite{chentsov} experiments
considered the time dependent vector potential associated with a laser beam
(i.e. a coherent, focused electromagnetic wave). Although the laser system considered in \cite{chiao, ageev, chentsov}
is different from the time dependent solenoid considered here, it is easy to see from
the proposed covariant phase shift in terms of the Faraday 2-form as given in \eqref{cov-2}
that the magnetic field part coming from ${\bf B} = \nabla \times {\bf A}$ will always
cancel the electric field part coming from ${\bf E} = - \partial _t {\bf A}$. There could still
be phase shift coming ${\bf E} = -\nabla \phi$ if there is a non-zero scalar potential in
addition to the time varying vector potentials (for both the time varying solenoid and the
laser system considered in \cite{chiao, ageev, chentsov} $\phi=0$). Now, the one experiment
we did find to actually test the time dependent AB effect did find {\it no
time varying phase shift} \cite{chentsov} which would then favor our
proposed expressions for the covariant AB phase shift \eqref{cov-1} or \eqref{cov-2}.
It should be stressed that some of the authors of the experiment \cite{chentsov}
argued that their null result was due to inadequacies of the experiment \cite{ageev}.
Further experiments are needed to confirm which prediction is correct.

From the above explicit calculation from equations \eqref{A} -- \eqref{ephase} of the cancellation
of the standard AB phase \eqref{Aint-2} with the phase contribution coming from the electric field
\eqref{ephase} one can surmise that in order to see this effect (or rather non-effect since the two
contributions are predicted to cancel) there needs to be some conditions or relationship between the
time scale of the variation of the magnetic field, $t_B$, with respect to the time-scale of the electron
to travel from the source to the screen, $t_{electron}$. If one assumes that the magnetic field varies
sinusoidally as in \cite{chiao, ageev, chentsov} with a frequency $f _B$ then $t_B = \frac{1}{f _B}$.
Assuming that the electrons move with velocity $v_e$ and if the distance between the screen and the
electron source is $L$ the time scale of the electrons is $t_{electron} = \frac{L}{v_e}$. Thus, to see this
non-effect one needs $t_{electron} \sim t_B =  \frac{1}{f _B}$. If one has
$t_{electron} \ll t_B =  \frac{1}{f _B}$ -- the time scale of the electrons is much less that the time
scale of the magnetic field variation -- then one will get the phase shift of the static situation since the
electrons move through the field much faster than it changes so that the field is effectively static.
If, on the other hand, $t_{electron} \gg t_B =  \frac{1}{f _B}$ -- the time scale of the electrons is much
greater than the time scale of the magnetic field variation -- then the effect of the magnetic
field on the phase shift of the electron will time average to zero. For the set-up in \cite{ageev, chentsov}
and the proposed experiment in \cite{chiao} the speed of the electrons was $v_e \sim 10^7 \frac{m}{s}$.
Assuming $L \sim 0.1 ~ m$, one gets $t_e \sim 10 ^{-8} ~ sec$. Thus, to see this cancellation
of the standard AB phase with the phase coming from the electric field ${\bf E} =-\partial _t {\bf A}$,
one needs the magnetic field to vary on a time scale of $t_B \sim 10^{-8} ~ sec$ or change
with a frequency of $f _B \sim 10^8 ~ Hz$. If one wanted to have a higher/lower frequency,
one should adjust the velocity of the electrons to be higher/lower according to the
relationship $v_e \sim f_B L$, e.g. for $f _B \sim 10^3 ~ Hz$
and with $L \sim 0.1 ~ m$, one should take $v_e \sim 10^2 \frac{m}{s}$. \\
%
%
\par\noindent {\bf 4. Conclusions} \\
%
%
\par
One of the most important phenomena which lies at the interface
of gauge theories and quantum mechanics is the Aharonov-Bohm
effect \cite{AB,ES} - an extra phase shift in the
interference pattern of the quantum mechanical double-slit experiment
due to the presence of electromagnetic vector, ${\bf A}$, and
scalar, $\phi$, potentials. The expressions for the magnetic and
electric AB phase differences are given by \eqref{phase-a1} and
\eqref{phase-p1}, respectively. These expressions are
non-covariant and thus one can ask for a covariant expression
which should combine/unify \eqref{phase-a1} and \eqref{phase-p1}.
In this paper, we have examined two possible covariant
generalizations, namely equation \eqref{cov-1} and \eqref{cov-2},
of the non-covariant electric and magnetic phase
differences. Expression \eqref{cov-1} was in terms of the space-time
line integral of the four-vector potential,
and expression \eqref{cov-2} was in terms of a space-time surface area integral
of the Faraday 2-form. Both expressions reduce to the non-covariant AB phase
differences \eqref{phase-a1} and \eqref{phase-p1} in the static limit.
Additionally, for the time dependent case of an infinite solenoid both \eqref{cov-1} and \eqref{cov-2}
gave the same, somewhat surprising result that there would be no
time dependent AB phase shift. One would only have whatever static AB-phase shift
existed before the start of any time variation of the magnetic flux. For the expression \eqref{cov-1}
in terms of the space-time line integral of the four-vector potential this null result
was the result of the cancellation between a true AB phase shift, i.e. expression \eqref{Aint-2},
and a non-AB type phase shift \eqref{ephase} due to the electric field that exists outside
the solenoid in this case. Since the electron for the time dependent, infinite solenoid did not move
in a field-free region (and in addition the force on the electron was not zero) this
is some generalized, or hybrid Aharonov-Bohm effect with part of the shift
coming from the potential and the other part coming from the fields.

The expression for the AB phase difference given in terms of the
Faraday 2-form, \eqref{cov-2}, shows that the cancellation
of the magnetic part of the AB phase coming from $\nabla \times {\bf A}$
will generally be canceled by the $-\partial _t {\bf A}$ part of the electric part
of the AB phase. For the time dependent case, this then leaves only the part of the electric
AB phase coming from the scalar potential $\phi$. For the case of a time-varying
magnetic field inside an infinite solenoid (or for the laser set-up considered
in \cite{chiao, ageev, chentsov}), $\phi =0$, and
thus one gets no time varying AB phase difference.

As a final comment, we note that the expression for the AB phase
difference in \eqref{cov-1} is essentially a Wilson loop
\cite{wilson} which is used to study the issue of confining versus
non-confining phases, i.e. the ``area" law versus ``perimeter" law,
of Yang-Mills gauge theories like QCD. In this work, we are making the suggestion of
the equivalence of the ``perimeter"
integral in \eqref{cov-1} with the ``area" integral in
\eqref{cov-2}. Thus, it maybe of interest to study non-Abelian
gauge fields using the proposed AB phase difference given in \eqref{cov-2}. \\
%
%
\par\noindent {\bf Note Added}
%
%
\par\noindent
After this paper was accepted for publication, we learned of the related work of
\cite{rousseaux} and \cite{moulopoulos} dealing with similar issues and having some of the 
same conclusions. \\
%
%
\par\noindent {\bf Acknowledgments}
%
%
\par\noindent
We thank Mihalis Dafermos and Atsushi Yoshida for useful correspondences.
The work of DS was supported via a 2012-2013 Fulbright Senior
Scholars Grant.
%
%

\end{document}